# Impact of Resin Molecular Weight on Drying Kinetics and Sag of Coatings

*Marola W. Issa[1], Steve V. Barancyk[2], Reza M. Rock[2], James F. Gilchrist[3], and Christopher L. Wirth[1]*

[1]Department of Chemical and Biomolecular Engineering, Case School of Engineering, Case Western Reserve University, Cleveland, Ohio 44106, United States

[2]PPG Industries, Inc. Pittsburgh, Pennsylvania 15272, United States

[3]Department of Chemical and Biomolecular Engineering, Lehigh University, Bethlehem, PA 18015, United States

**Corresponding author**

Christopher L. Wirth

Chemical and Biomolecular Engineering Department

Case School of Engineering

Case Western Reserve University

Cleveland, OH 44106

wirth@case.edu

MWI Orchid: 0000-0002-8099-0459

CLW Orchid: 0000-0003-3380-2029

**Abstract**

Coating performance is influenced by factors inherent to formulation design and processing conditions. Understanding the complex interplay of these attributes enables for mitigating defects, such as sag, which is a gravity-driven phenomenon that impacts the aesthetic and functionality of a coating. The work herein investigates the impact of resin molecular weight and solvent choice on drying kinetics and sag velocity in polymer films. These films, ranging in thickness from ~60 µm to ~120 µm were formulated with 45% by weight polymer resin in one of two solvent packages with different relative evaporation rates (RER). Gravimetry was initially used to track drying kinetics and a one-dimensional diffusion model was utilized to compute the apparent solvent diffusivity. In addition, the film thickness was tracked with optical profilometry. Results from these measurements showed that for fixed molecular weight the drying kinetics increased by approximately two-fold for the high RER solvent, whereas the apparent diffusivity tended to increase with increasing polymer molecular weight. Films formulated from higher molecular weight resins had greater initial viscosities and thicknesses for identical draw down blade clearance. By extension, the higher apparent diffusivities at greater molecular weights were attributed to effects of prolonged evaporation times for the thicker films. The sag velocity was measured through the thickness of the film for these systems at a 5° incline using the Variable Angle Inspection Microscope (VAIM). Measurements showed an increase in sag velocity for thinner and less viscous films, which was somewhat surprising both because a thinner film will experience lower gravitational stress and quicker drying times as compared to a thicker film. From these data we conclude that formulating a coating with higher molecular weight resin, although likely to increase drying time, will tend to deter sag because of the large impact of viscosity on these phenomena.

1. **Introduction**

Coatings are a ubiquitous technology, extending from surfaces we see and touch to those we consume (i.e., pharmaceutical coatings)[1–5]. Coatings serve various purposes, contributing to both functionality and durability while enhancing aesthetics[6–8]. These complex, multi-component systems are formulated with a precise balance of materials to address specific application needs. For instance, many coatings provide essential protection, contributing to the longevity of the underlying substrates[9]. As an example, coatings are used for offshore oil and gas operation facilities, marine pipeline, ships, and bridges to minimize corrosion, in addition to meeting a broad range of application-specific requirements such as fire protection, long-term durability, and reduction of biofouling[10–13]. Automotive coatings are specifically formulated to include ultraviolet light-resistive and water-repellent components ensuring durability and aesthetic appearance[8]. Additionally, coatings can impart diverse functionalities to a surface, contributing to properties such as its electrical conductivity, catalytic activity, and wetting properties[14–18].

Efforts to advance formulations are primarily motivated by performance improvement, enhanced functionality, new manufacturing processes, or sustainability goals[19]. Notably, improving the sustainability of formulations, for example through reduction of volatile organic compounds (VOCs), is increasingly important to corporate entities and policy makers[20,21]. Given the complexity of a typical coating formulation and the key role that VOCs and other substances of concern may play in mediating coating properties, one common challenge associated with engineering new formulations to meet these needs is that of defect mitigation. Defects arise due to various factors and often materialize during the solvent flash stage. This post-application phase precedes the curing of the coating through UV, IR, or heat exposure. Local nonuniformities in the coating and surface tension gradients can lead to various defects[22,23], including orange peel and crater formation. Another common defect is coating sag, which is characterized by a gravity-driven flow on vertical or inclined substrates resulting in an uneven surface[24]. Potential causes include excessive film thickness, insufficient degree of atomization during spray application resulting in large droplets and higher degrees of solvent retention, or poor choice of solvent. Sag not only compromises the aesthetic functionality of the finish, but also demands re-work in manufacturing lines, which can be energy and time intensive[25,26].

Minimizing sag defects requires a comprehensive understanding of the various factors at play. Parameters inherent to formulation design such as viscosity and solvent choice, as well as processing conditions including substrate geometry and film thickness, work together to impact performance[8]. Automotive clear coats typically consist of three essential components, namely, solvents, binders, and additives[27]. Immediately following application, the solvents begin to evaporate. Evaporation is influenced by the application conditions and the physicochemical properties of the polymeric coating[28]. Further, drying kinetics are governed by two fundamental processes: solvent volatility and diffusive transport through the film[29]. Recognizing the importance of formulation choice and its impact on drying kinetics, various studies have explored solvent diffusion, emphasizing mass and heat transfer[30–33]. In polymer-solvent systems, solvent diffusivity was seen to decrease by several orders of magnitude as the solvent concentration decreases[34–36]. Further, in a multicomponent system, preferential loss of a solvent can result in the polymer being left in a system with an incompatible solvent. This could lead to polymer precipitation and coating failure. Earlier work concluded that for lower polymer concentrations, the diffusion coefficient changes significantly with the molecular weight, whereas at higher polymer concentrations, the diffusion coefficient was indistinguishable and independent of the molecular weight[37]. Several models were developed to help interpret drying kinetics. Solvent diffusivity in binary solutions of varying polymer concentrations were understood using free volume theory[38,39], Flory Huggins theory, kinetic theory[40], and entanglement theory[41]. These studies concluded that the diffusion coefficient decreases with increasing polymer concentration. Also, the solvent self-diffusion coefficient decreases with increasing the molecular size of the solvent[29,41]. Further, the effects of polymer molecular weight and interactions between the polymer and solvent on film thickness and topography were investigated. The studies showed that increasing polymer molecular weight results in thicker films during spin coating[42]. Additionally, there are several studies suggesting that the transport properties and solvent diffusion in nanofilms are dependent on the thickness and substrate chemical structure[43–45]. Further, increased polymer concentration leads to reduced chain mobility and solvent evaporation[39]. These studies offer in-depth insights into the impact of coating properties, and drying conditions on solvent diffusion.

Given the importance of minimizing sag to improve coating operation efficiency and quality, a variety of optical and non-optical methods have been developed to quantify this phenomenon, including the indicator band method, the falling wave technique, and particle-based microscopy techniques[24,46–50]. Further, sag evaluation in industry relies on a visual technique developed by the American Society of Testing and Materials (ASTM D4400). This method, which employs an anti-sag meter, generates a series of parallel films with variable thickness. Measurements provide an index value corresponding to the largest thickness at which paint lines do not merge due to sag. While these measurement techniques have practical importance for quickly testing many different formulations, it is difficult to obtain physicochemical understanding of the sag process because of their simplistic and qualitative nature. To address these limitations, we recently developed a particle-based microscopy technique called Variable Angle Inspection Microscopy (VAIM), designed for real-time quantitative measurements of the transient flow-fields in films (see **Section 3**).

The aim of this work is to experimentally investigate the impact of formulation properties and processing conditions on sag in drying coatings. The effort integrates the study of drying kinetics, influenced by resin molecular weight and application conditions, with real-time sag evaluation via VAIM. The systems described herein consist of model solvent borne liquid automotive coatings that form a soft solid via evaporation. These model clearcoat solutions, composed of 45% polymer by weight, behaved as Newtonian fluids over the range of shear rates relevant to this study before evaporation commenced. The formulation of each coating was systematically adjusted via resin molecular weight and by incorporating high and low Relative Evaporation Rate (RER) solvents. Drying kinetics was evaluated via gravimetry. A one-dimensional diffusion model, with dynamic thickness measurements, was utilized to compute the solvent diffusion coefficients. The model assumes constant diffusion, which is a reasonable approximation for short intervals[51]. We further measure the velocity distributions in such films and assess sag as a function of resin molecular weight. This allowed for investigating the molecular weight dependence of diffusivity, providing insights into sag performance in drying automotive coatings. Our results suggested competing contributions of viscosity and processing lead to a complex sag response.

## 2. Theory

### 2.1. Solvent Diffusivity.
Solvent diffusivity was obtained from experimental measurements by applying an existing model for transport in a plane sheet. The model is based on the one-dimensional equation for mass diffusion in a film with a thickness $L(t)$[51] (**eq. 1**).

$$\frac{\partial C(x,t)}{\partial t} = D \frac{\partial^2 C(x,t)}{\partial x^2}, \text{ where } 0 \leq x \leq L(t) \tag{1}$$

Where $C(x,t)$ is the solvent concentration as a function of position and time and $D$ is the solvent diffusion coefficient. The solution assumes constant diffusion at early drying times, negligible advection for a quasi-stationary phase, uniform initial concentration, $C_0$, and a boundary condition such that the substrate is impermeable (zero solvent flux through the solid interface), $\frac{\partial C(0,t)}{\partial x} = 0$. The second boundary condition is set by the mass flux, J, which is proportional to the concentration gradient at the upper surface, $J = \rho_v D_v \frac{\partial C(L,t)}{\partial x} = k(c(L,t) - c_\infty)$ and is also assumed proportional to the concentration difference between the surface and the bulk vapor phase, where $\rho_v$ is the density of solvent vapor and $k$ is a proportionality constant. The model assumes no transport resistance in the vapor phase and that the atmospheric concentration, $c_\infty$, is zero. Solving via these boundary conditions yields the following expression (**eq. 2**).

$$\frac{C_0 - C}{C_0} = 1 - \frac{4}{\pi} \sum_{n=0}^{\infty} \frac{(-1)^n}{2n+1} \exp\left\{-D(2n+1)^2 \pi^2 \frac{t}{4L(t)^2}\right\} \cos\frac{(2n+1)\pi x}{2L(t)} \tag{2}$$

Evaluating the volume integral of the above mathematical equality for thickness values ranging between 0 and $L(t)$ results in an expression of the normalized amount of solvent that has evaporated from the polymer film with respect to the total amount present initially (**eqs. 3, 4, 5**).

$$\int \frac{C_0 - C}{C_0} dV = \int_0^{L(t)} \left[1 - \frac{4}{\pi} \sum_{n=0}^{\infty} \frac{(-1)^n}{2n+1} \exp\left\{-\frac{D\pi^2 t(2n+1)^2}{4L(t)^2}\right\} \cos\frac{(2n+1)\pi x}{2L(t)}\right] y\, z\, dx \tag{3}$$

$$\frac{M_t/C_0}{L(t)\, y\, z} = \left[x - \frac{4}{\pi} \sum_{n=0}^{\infty} \frac{(-1)^n}{2n+1} \exp\left\{-\frac{D\pi^2 t(2n+1)^2}{4L(t)^2}\right\} \frac{2}{(2n+1)\pi} \sin\frac{(2n+1)\pi x}{2L(t)}\right]_0^{L(t)} \tag{4}$$

$$\frac{M_t}{M_\infty} = 1 - \sum_{n=0}^{\infty} \frac{8}{(2n+1)^2 \pi^2} \exp\left\{-D(2n+1)^2 \pi^2 \frac{t}{4\,L(t)^2}\right\} \qquad (5)$$

Where the volume is equal to the product of thickness, $L(t)$, width, $y$, and film length, $z$. $M_t$ is the amount of solvent that had evaporated at a given time, and $M_\infty$ is the total amount of solvent that can be lost at the end of the drying process. The model described above requires an instantaneous measure of film thickness $L(t)$ to furnish predictions. The present expression was modified to capture changes in film thickness within a finite liquid volume, accounting for leveling and evaporation effects. Further, film thinning was measured and subsequently incorporated into the desorption model to enhance the goodness of fit. As will be described in detail in the experimental section, the mean film thickness for each scan was evaluated at a distance from film edge on the order of one capillary length, $l_{cap} = (\sigma/g\rho)^{1/2}$, $\sigma$ is surface tension and $\rho$ the fluid density. The capillary length was approximately ~ 1 mm. Lastly, the desorption model was evaluated for seven summation terms (i.e., n = 0 to n = 6) based on the plateauing of the value of diffusivity (see **Fig. S1** in Supplemental Information).

**2.2.  Drying Kinetics and Sag of Polymer Films.** Liquid coatings during the flash stage exhibit evaporation-induced drying. This phenomenon occurs as the solvent diffuses out through the polymer matrix, increasing polymer concentration and decreasing in film thickness. Experiments were conducted to track the mass loss over the first 5000 seconds of flash. Shortly after film casting, a roughly constant drying rate was observed. The rapid decrease in coating mass was followed by a gradual tapering off around the ~ 2-hour mark. The coating mass is observed to have an exponential decrease with time, thus fit by **eq. 6**. The time constant associated with the exponential function, $\alpha$, was then used to compare formulations of varying molecular weights and solvent volatilities.

$$M(t) = A * e^{-\alpha t} + M_0 \qquad (6)$$

Where $M(t)$ is the sample mass at time $t$. $A$ is the mass of material undergoing evaporation, $M_0$ is the offset of the exponential curve, which represents the material that does not evaporate, and $\alpha$ is the exponential rate constant in dimensions of inverse time. Although experimental data were fit with **equation 6** allowing all adjustable parameters $\alpha$, $A$, and $M_0$ to float, note the total mass at t = 0 ($M(0) = A + M_0$) was typically within 1% - 0.1% of the experimental value. Although the mass of material that will not evaporate is known *a priori*, $M_0$ may include trapped solvent at the end of the experiment.

For steady, one-dimensional flow down an incline, liquid films experience competing effects of gravitational and viscous forces. Gravity works to pull the liquid downstream, whereas viscosity opposes flow. The maximum velocity is at the coating free surface, with subsequent decrease in magnitude as the region of interest (ROI) approaches the substrate (see **Fig. S2** in Supplemental Information). The velocity flow field is influenced by competing effects of material properties (i.e., viscosity and density) and film thickness[1]. Further, we expect from previous work conducting blade application of polymer films that formulation viscosity impacts the applied film thickness[52]. The data collectively suggests that for coatings with the same initial thickness, $L(t)_1 = L(t)_2$, an increase in viscosity leads to decreased average velocity. Though, the application conditions herein were kept constant, resulting in viscosity-dependent film thickness such that higher viscosity samples resulted in thicker films. Thus, the sag velocity measured across formulations exhibits variable viscosity and thickness contributions, resulting in a complex sag response.

## 3. Materials and Experimental Methods

**3.1. Materials.** Fluorescent silica particles (~ 1 μm) were synthesized through the Stober method[53]. A detailed description of this process was highlighted in previous work[1]. Briefly, tetraethyl orthosilicate (TEOS) was combined with a mixture of ethanol (200 Proof, 99.5% purity) and ammonium hydroxide ($NH_4OH$). This initiates a hydrolysis and condensation polymerization reaction which generates monodispersed silica spheres. Rhodamine-B Isocyanate (RBTC) was added to impart fluorescence. Formulations containing probe particles (~ 0.1 % by volume) were transferred to a bath sonicator to ensure homogenous mixing.

Twelve acrylic-based model clear coats with known polymer molecular weights, polydispersity index, and compositions were utilized. The polymers are based on different ratios of acrylic, methacrylic, and styrene monomers that have different glass transition temperatures $T_g$. All twelve polymers were free-radical polymerized with a PPG-standard procedure, producing statistical copolymers with linear architecture. The molecular weights were measured with a gel permeation chromatographer with tetrahydrofuran (THF) as the eluent at room temperature. Reported values are weight averaged molecular weights ($M_w$) calculated relative to a polystyrene standard. They will differ from absolute values depending on the structural chemical differences between each statistical copolymer and the polystyrene standard. The model clear coats were composed of 45% resin by weight and 55% solvent mixtures (PPG Industries), consistent with established standards for automotive formulations. The resin molecular weights varied from 1744 to 16735 g/mol (see **Table 1**). The copolymer resins were formulated to have calculated Fox $T_g$ of 0 or 30 °C, below and above the experimental temperature, by pairing low, medium, or high $M_w$.

The solvents used in this work consisted of mixtures with high relative evaporation rate (RER = 87, Butyl Acetate = 100) and low relative evaporation rate (RER = 27). The corresponding vapor pressures were calculated to be 0.739 Kpa and 0.252 Kpa for the High and Low RER solvent mixtures respectively[54]. The formulations were prepared by transferring the polymer-solvent suspensions in tightly sealed glass scintillation vials and placing them on a roller mixer (Benchmark Scientific R3005 Tube Roller). Further, samples were allowed to gently mix for a few hours and were then left to sit overnight before proceeding with measurements.

**3.2.  Coating Preparation and Characterization**. Film application was performed with an automatic drawdown machine (Byko Drive G. 2122), 3-inch adjustable blade coater, and a custom-made plate designed to hold a 25 X 75 mm microscope slide. A set volume of polymer solution (80 µL) was carefully dispensed on the microscope slide via a micropipette. The automatic film applicator, positioned at a zero-incline angle, was then set to motion at velocity of 25 mm/s. The casting knife gap thickness was set at 254 µm application thickness, with the wet film thickness varying with the polymer molecular weight. Following application, the samples were immediately transferred to an automatic microbalance, which auto-logged the sample mass at 3-second increments for two hours. Five to six experimental repeats were carried out for each formulation at the laboratory temperature of $\sim 22 \pm 1$ °C.

Further, the film thickness was measured via an optical surface topography apparatus NANOVEA ST400. The profilometer has a lateral resolution of 2.6 µm and a 1.1 µm depth-range. Samples were placed at 0° and 5° incline planes and were allowed to dry at the laboratory conditions. The film thickness was measured at a rate of one scan per minute for a total of 10 minutes corresponding to flash stage. The scans were conducted at the center of the film along the direction of the blade during application of the coating. The mean thickness was evaluated with a MATLAB script and an average of three experimental repeats was utilized in the desorption model.

This allowed for examining the effect of solvent evaporation and gravitational force on the apparent film thickness.

Rheological characterization of the model clear coats was carried out via a stress-controlled rheometer (HAAKE MARS III). The measurements utilized a cone-plate geometry (diameter: 35 mm, cone angle: 1°) and a solvent trap to minimize solvent evaporation. The samples were sheared at 100 s$^{-1}$ for 200 sec and the shear rate was then varied and reduced to 0.1 s$^{-1}$. Further, solution densities were measured via a density meter (Anton Paar DMA$^{TM}$ 4500 M). Tabulated data, measured at 22 °C, are provided in the results section.

**3.3. Video Capture and Particle Tracking**. Sag was characterized via VAIM with epi-fluorescence measurements of tracer particles dispersed in the model coatings with sufficiently low number concentration to not induce deviations from Newtonian rheology (data not shown). VAIM was previously benchmarked and described[1], but briefly, the main components of this microscope are a CMOS digital camera (Thorlabs DCC1240C-HQ), 20x objective (Olympus UPLFLN20X), tube lens, light source (white light), fluorescence filters (Thorlabs TXRED MF559-34), and a goniometer which enables for positioning the VAIM at any desired angle. The fluorescence components include a white light input, excitation and emission filters, and a dichroic filter. Emission (630 ± 34.5 nm) and excitation (559 ± 17 nm) filtering was achieved with a Texas Red filer. Further, the setup includes a 3D-printed holder to secure the sample during measurements.

Sag measurements were conducted using five formulations containing high RER solvent and 45% polymer by weight. The samples exhibited viscosities ranging from 0.026 – 0.11 Pa s, allowing the assessment of sag performance across a range of viscosities. Three individual runs were performed for each formulation. Immediately after film application, the coated substrates were transferred to the VAIM, positioned at 5° angle, for sag measurements that consisted of fluid flow in response to gravitational stress and solvent evaporation (see **Fig. 1**). Experiments were conducted at a laboratory temperature of ~ 22 ± 1 °C. As was done in previous work, the scan path for the objective was controlled with a piezo and defined based on film thinning data from optical profilometry[1]. This approach enabled the path of the objective to maximize the number of measurements between the dynamic boundary (i.e., free surface) and substrate. Although the initial film thickness varied with viscosity, the scanning process followed the same steps. Namely, a given ROI was observed for 5 seconds and the piezo then moved to a different vertical position located 10 µm away. The objective was scanned in an oscillatory pattern, which was set to start at the free surface, go down to the substrate, and then back up to the free surface. Hence, the measured velocity displays a distinct sawtooth shape (see **Fig. S2** in Supplemental Information).

RGB color images with dimensions of 1280 x 1024 pixels were collected during the experiments. Data was converted to an 8-bit format and the dimensions were reduced to 900 x 800 pixels. This allowed for including most of the captured tracer particles. All videos were collected at a rate of 10 frames per second (FPS) with exposure time of 70 ms. Videos were analyzed with a particle tracking algorithm executed in MATLAB and the average velocity in each plane was recorded[55].

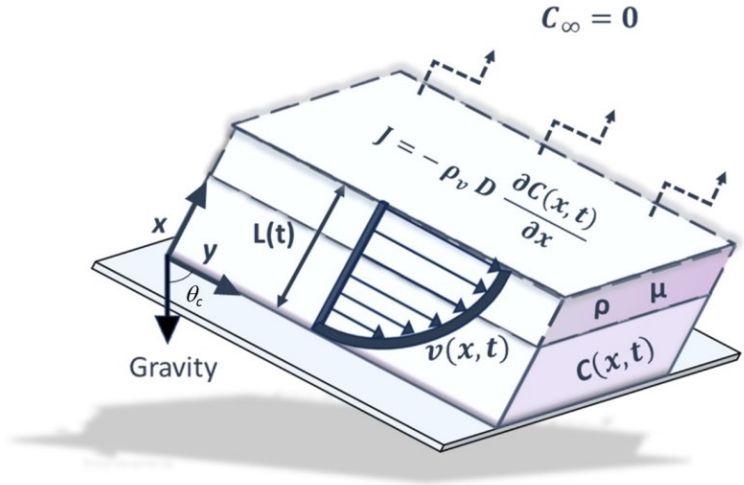

**Figure 1: Schematic of an evaporating film with known initial viscosity and thickness.** The VAIM was positioned at an inclined angle, θ, allowing for velocity measurements as the coating flows. The piezoelectric objective scans the various depths of a falling liquid film and tracks the local velocity as the film flows and dries. Note the angle $\theta_c$ is the complement to the angle of inclination reported herein, $\theta = 90 - \theta_c$.

## 4. Results and Discussion

**4.1. Characterizing the Model Clear Coat Systems.** Rheological and density measurements were conducted with the aim to better understand how physicochemical features of the model clear coats varied. **Figure 2(a)** displays the viscosity flow curves as a function of shear rate for the model clear coats in the absence of solvent evaporation. The figure shows samples with varying molecular weights, formulated with 45% polymer by weight and a high RER solvent. The shear rates selected for rheology measurements aimed to simulate the shear experienced by the coating during sag (~ 1 s$^{-1}$). As can be seen from these data, the model clear coats exhibited Newtonian behavior for shear rates ranging between 0.1 s$^{-1}$ and 100 s$^{-1}$. In addition, model clear coat samples containing tracer particles were also measured and displayed no change in the average viscosity suggesting that the addition of particles had no discernible impact on bulk rheology. As expected, the average viscosity increased with increasing the polymer molecular weight (see **Figure 2(b)**). Larger polymer chains with higher molecular weights tend to increase viscosity and resistance to flow[56]. Notably, the increase in viscosity may not be linear, as it depends on several factors including entanglement and intermolecular forces[57,58]. We anticipate that viscosity will play an important role in the processing conditions explored herein, including film thickness and wettability. Further, low viscosity enhances capillary-induced spreading, while increasing viscosity is one strategy for preventing sag[49].

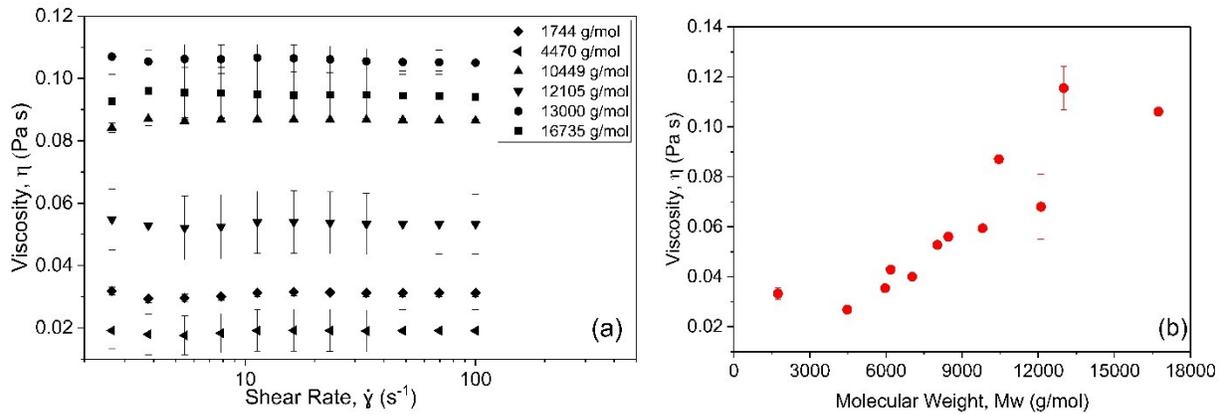

**Figure 2: Rheological behavior of model clear coats.** The viscosity flow curves (Pa s) as a function of (a) shear rate (s$^{-1}$) for samples with varying molecular weights formulated with a high RER solvent and 45% polymer by weight. The error bars represent the standard error obtained from three experimental repeats. These measurements suggest a Newtonian response in the sag regime probed herein. (b) The average shear viscosity, obtained from three experimental repeats. Viscosity increased with increasing polymer molecular weight.

Formulations were further characterized by measuring the density of the model clear coats. Previous work has shown that density can significantly impact diffusivity and drying kinetics[59]. **Table 1** summarizes the densities and their corresponding polymer volume fractions, φ. The computed volume fractions based on the measured densities and known molecular weights were observed to be nearly identical but displayed a very slight decrease with increasing the polymer molecular weight. These values were ~ 42% v/v and ~39% v/v for the low and high RER systems respectively.

| Formulation No. | Molecular Weight (Mw) g/mol | Polydispersity Index (PDI) | Solvent ID | Average Density g/cm$^3$ | Polymer Volume Fraction φ | Solvent ID | Average Density g/cm$^3$ | Polymer Volume Fraction φ |
|---|---|---|---|---|---|---|---|---|
| 1 | 4470 | 2.3 |  | 0.992 | 0.423 |  | 0.964 | 0.394 |
| 2 | 10449 | 3.2 |  | 0.994 | 0.422 |  | 0.966 | 0.392 |
| 3 | 13000 | 3.1 |  | 0.995 | 0.421 |  | 0.966 | 0.392 |
| 4 | 1744 | 1.3 |  | 0.994 | 0.422 |  | 0.965 | 0.393 |
| 5 | 12105 | 3.6 |  | 0.996 | 0.421 |  | 0.968 | 0.391 |
| 6 | 16735 | 4.9 | Solvent (RER 27) | 0.995 | 0.421 | Solvent (RER 87) | 0.967 | 0.392 |
| 7 | 5965 | 2.6 |  | 0.993 | 0.423 |  | 0.959 | 0.397 |
| 8 | 7032 | 2.8 |  | 0.993 | 0.423 |  | 0.969 | 0.391 |
| 9 | 8456 | 3.0 |  | 0.993 | 0.422 |  | 0.963 | 0.395 |
| 10 | 8023 | 3.1 |  | 0.996 | 0.421 |  | 0.968 | 0.392 |
| 11 | 9803 | 3.3 |  | 0.996 | 0.421 |  | 0.968 | 0.391 |
| 12 | 6184 | 2.8 |  | 0.995 | 0.421 |  | 0.967 | 0.392 |
| Pure Solvent (RER 100) | 116.2 | – | – | 0.995 | – | – | – | – |
| Pure Solvent (RER 12) | 146.2 | – | – | 0.875 | – | – | – | – |

**Table 1:** Physical properties of the polymer-solvent formulations at 21°C

**4.2. Drying Kinetics.** As described in preceding sections, the exponential rate constant for the experimental systems was obtained from gravimetry by measuring mass as a function of time while the solvent evaporated. During the main period of evaporation the sample mass displayed an exponential decrease that was subsequently fit (see **eq. 6**) to obtain a value of $\alpha$ for each combination of resin and solvent. Representative data, consisting of five repeats of a coating with a resin molecular weight of 4470 g/mol, are shown in **Figure 3**.

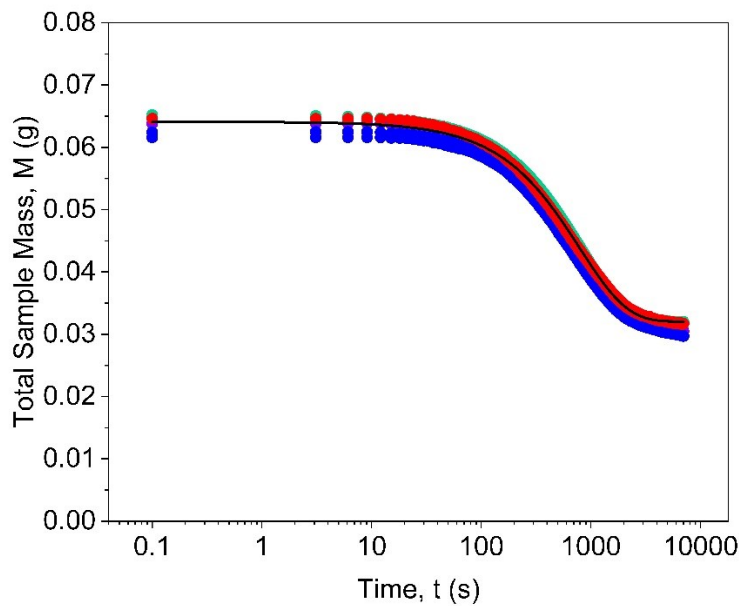

**Figure 3: Gravimetry plot for films with a 4470 g/mol polymer molecular weight and a high RER solvent.** Data shown here represent five experimental repeats. The x-axis is the observation time (seconds), and the y-axis is the sample mass (grams). Data collection was automated with the microbalance to capture data at 3-second intervals over 5000 seconds. Mass loss kinetics displayed an exponential decrease for all samples of varying polymer molecular weights. The rapid decrease in coating mass was followed by a gradual tapering off around the ~ 2-hour mark where a second plateau is reached. This value amounts to ~ 95% solvent loss for these systems indicating the presence of ~5% residual solvents in the film. The percentage of residual solvents displayed a strong dependence on the resin molecular weight and solvent volatility in early drying times. However, in the later stages of drying, the rate exhibits strong molecular weight dependence (i.e., viscosity) and weak dependence on volatility. The black solid line corresponds to a fitted curve from one experimental run.

Residual solvent content $M_{prs}$ was defined as the percentage of solvent from the initial formulation that was retained in the coating at a given time point. Herein, we analyze this quantity at both short (~10 minutes) and long (~2 hrs) times (see **Fig. S5 & S6** in Supplemental Information). Analysis of the residual solvent content at short times corresponding to the flash stage showed different retention percentages for low and high molecular weight resins in both high and low RER. Coatings formulated with high RER solvents displayed retention percentages of 45% and 65% for low and high molecular weight resins, respectively. Further, solvent retention increased for low RER solvents with percentages ranging from 68% for low molecular weight resin to 80% for high molecular weight resins (see **Fig. S5** in Supplemental Information). These initial results suggest the drying kinetics are a strong function of both the resin molecular weight and vapor pressure of the solvent. Increasing the resin molecular weight tended to increase retained solvent at short times consistent with the flash stage, whereas increasing the vapor pressure of the solvent tended to decrease solvent retention at short times relevant to the flash stage.

Interestingly, although the impact of resin molecular weight on retained solvent remained at long times (~2 hrs), the effect of solvent volatility did not. At the end of drying, approximately 5% and 14% of solvent was trapped in the coating for the limits of low and high molecular weight resins, respectively (see **Fig. S6** in Supplemental Information). However, the solvent content trapped exhibited similar values, irrespective of the solvent's RER at these longer times. As the focus of this work is to assess the impact of formulation properties and processing conditions on sag performance during the flash stage, with subsequent thermal curing, we limited our investigation to the first 10 minutes. However, those such considerations may be relevant when considering the sustainable formulation and manufacturing of coatings. These data suggest the solvent retention levels beyond the initial flash stage primarily depend on the resin molecular weight and not the vapor pressure of the solvent. Other polymer properties could also play a role at these later stages, for instance the glass transition temperature $T_g$, the effects of which were not explored herein. Note that $T_g$ may primarily impact the later stages of drying such that initial solvent loss is independent of resin $T_g$[58]. Gravimetry data was further analyzed by obtaining the exponential decrease α for all formulations.

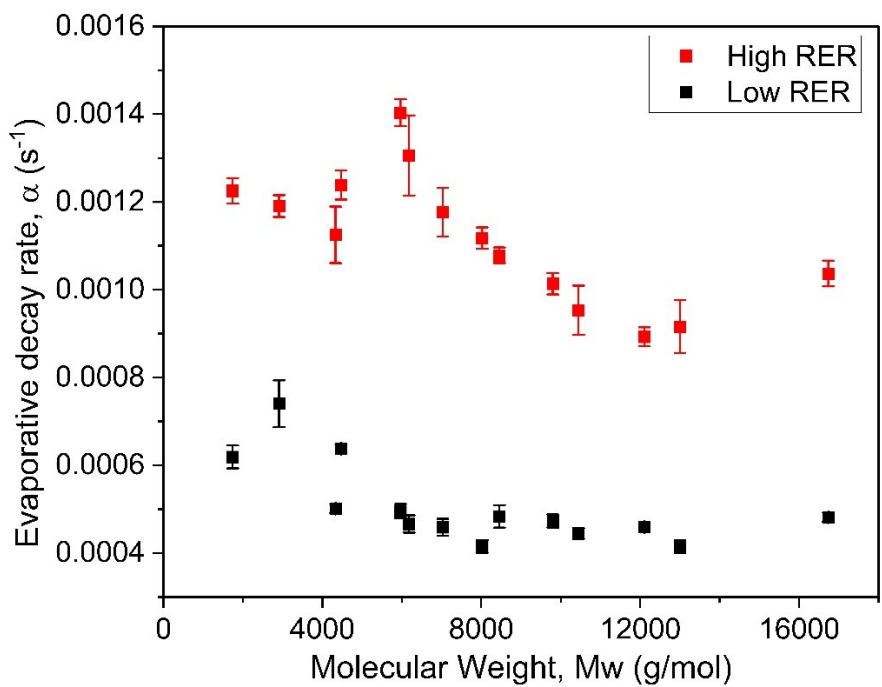

**Figure 4: Evaporative exponential rate constant, α (s⁻¹), as a function of polymer molecular weight (g/mol).** The black squares represent data obtained from formulation with a solvent characterized by a low relative evaporation rate. The red squares correspond to formulations that utilize a higher relative evaporation rate solvent. The error bars indicate the standard error calculated from six experimental repeats. These data suggest a dependence of rate constant on polymer molecular weight, demonstrating a decrease in rate as the molecular weight increases due to heightened viscosity. Formulations containing a higher RER solvent displayed a two-fold increase in the exponential rate constant.

**Figure 4** shows $\alpha$ as a function of polymer molecular weight. These initial data suggest that formulations containing higher polymer molecular weight dry at slower rates at fixed mass fraction. This trend was observed for both solvent systems with different values of RER. Further, coatings made with high RER solvent displayed a shift, approximately two-fold, corresponding to the vapor pressure of the solvent mixtures, which were calculated to be 0.739 Kpa and 0.252 Kpa for the High and Low RER solvent mixtures, respectively. The upward shift in data suggests faster drying kinetics with increased RER. This is in part attributed to the increased volatility, assuming the polymer solubility in the two solvent mixtures is similar[57]. The shift is also likely influenced by the density difference in the two solvent mixtures which was measured to be ~ 5%. Increased solvent molecular size and density leads to reduced diffusivity and drying rate[58]. Hence, the data shift can be attributed to these combined effects.

These data begin to suggest a physical picture of solvent evaporation during flash that is dependent on the unique combination of resin and solvent. As molecular weight increases or vapor pressure decreases, solvent tends to evaporate more slowly at short times (~10 minutes) corresponding to the flash step of a coating process. At longer times, the effect of solvent vapor pressure diminishes, but the impact of molecular weight remains such that higher molecular weight resin more effectively retains solvent. One would expect an increase in radius of gyration as molecular weight increases, which potentially could lead to an increase in polymer free volume in which the solvent can be retained[60,61]. The effect of solvent choice is more subtle. At short times, the higher RER solvent more quickly evaporates, increasing solids weight fraction and likely increasing the viscosity of the film. The rapid kinetics of drying at short times is then diminished such that later stages at long times are dominated by transport through a large viscosity fluid or semi-solid material, and desorption becomes a weak function of solvent volatility[61]. At this later stage, the physiochemical nature of the polymer itself, namely the molecular weight, primarily impacts this characteristic associated with drying.

**4.3. Film Thinning, Solvent Diffusivity, and Sag.** Film thickness of model clear coats was measured with optical profilometry during solvent evaporation and in response to a gravitational stress. The thickness was evaluated over an area near the center of the film on the order of one capillary length away from edges where a finite volume of film is uniform in thickness. In general, we expected the thickness of films to decrease over time considering the limited material volume and solvent evaporation. While the mean thickness exhibited an exponential decrease in magnitude, film edges tended to increase in thickness in time because of surface energy and the accumulation of solids because of the coffee ring effect[62,63]. **Figure 5** shows representative data of a full profilometer scan for a model clear coat formulation with 12105 g/mol $M_w$ polymer, positioned at 0°. These data show wet film thickness decreased from an initial value of 130 μm to around 50 μm exclusively because of solvent evaporation. Mostly pinned contact lines tended to enhance the coffee ring effect, as material was transported outward towards the edges. Interestingly, this phenomenon becomes more pronounced with increased material accumulation in the direction of drawdown. Note the progression of peaks upstream (left) with peak value ~ 70 μm and downstream (right) where it reaches ~ 150 μm.

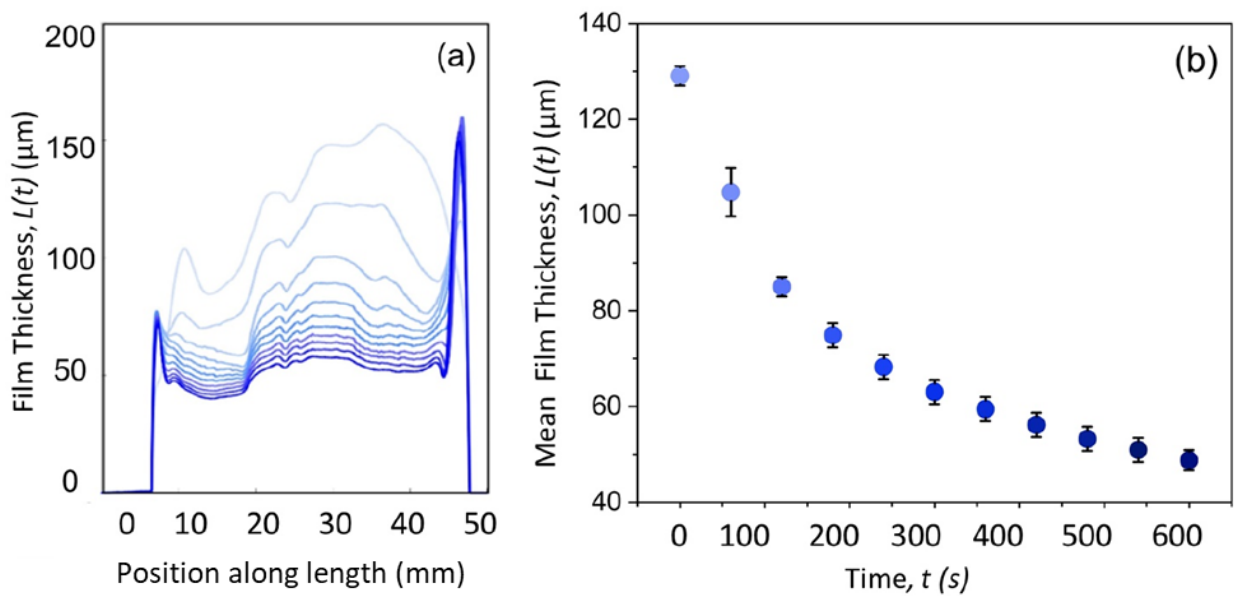

**Figure 5: Surface topography scans for formulations containing 12105 g/mol $M_w$ polymer, positioned at 0° incline.** The initial scan at time t = 0 is represented by the light blue line with an average thickness ~ 140 μm. Consecutive scans collected at 1 min interval are shown by color progression to dark blue at t = 10 min with average thickness ~ 55 μm. The x-axis depicts the position along the length (mm) where x = 0 represents the initial analyzed point with the scan direction in the positive x-direction. The drawdown direction is along the x-axis. As the solvent evaporates, material accumulation on the outer edge of the film becomes more pronounced. This is seen in the progression of peaks at the upper end of the scan with peak value ~ 70 μm and downstream where it reaches ~ 150 μm. Edge effects, on the order of the capillary length, were excluded from mean thickness measurements. Film mean thickness exhibits an exponential decrease as the solvent evaporates. Error bars represent the standard deviation associated with profilometry measurements for three experimental repeats. Note, similar profilometry scans were performed for coatings positioned at a 5° inclined plane. Those were utilized in the scan design to optimize sag measurements. Sag velocities were tracked in the region at ~ one-third of the distance from the coating upstream to minimize distortions in measurements.

The qualitative trends associated with the 'wet' film thickness described above were consistent across all measurements and formulations; note the 'wet' film thickness refers to the actual thickness of the film, whereas the drawdown clearance (of 10 mils or 254 μm) was identical for all films. **Figure 6(a)** displays the initial wet film thickness for the different polymer molecular weights. The applied film thickness was seen to increase with increasing the polymer molecular weight. This phenomenon was ascribed to effects of viscosity, which impacts the total mass flow rate during film application[52]. Often a simplified relationship is assumed where the film thickness is roughly half of the gap set by the blade. However, it is essential to recognize that variations in viscosity can significantly affect the applied film thickness and, consequently, the anti-sag predictions using the standard ASTM test. Herein, the viscosity-dependent thicknesses displayed an approximately linear trend with an increase in thickness as the viscosity increased (see **Fig. 6(b)**). Additionally, samples containing a low RER solvent displayed a slight increase in initial thickness as molecular weight of the resin increased. This is ascribed to the increased solvent viscosity which was measured to be ~ 0.01 cP and 1.3 cP for high and low RER solvents, respectively. **Figure 6(b)** displays the impact of viscosity on applied film thickness for the high RER model paints. Considering sag measurements utilize a fixed volume, we anticipate changes in thickness and viscosity to contribute greatly to flow performance.

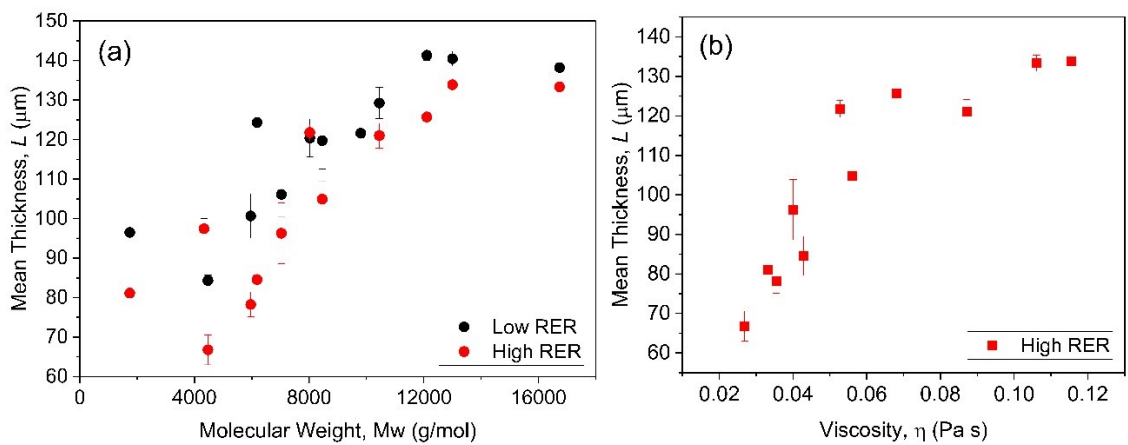

**Figure 6: Influence of molecular weight on wet film thickness.** (a) Initial mean wet thickness (μm) as a function of average polymer $M_w$ (g/mol). The black circles represent samples formulated with a low RER solvent. The red circles represent systems with a high RER solvent. The error bars denote the standard deviation obtained from three experimental repeats. Samples with low RER solvent exhibit a slight increase in initial mean thickness compared to those with high RER solvent, across different molecular weights. The shift in data was attributed to differences in solvent choice which contributes to the overall formulation viscosity. (b) The applied film thickness increased with increasing the molecular weight (i.e., viscosity). This is attributed to effects of viscosity which impacts the total mass flow rate and applied film thickness.

These data were used to determine the apparent solvent diffusivity from the one-dimensional desorption model described by the system of **eqs. (1) – (5)**. Diffusivity values were determined by measuring the normalized solvent loss over the total available solvent, $\frac{M_t}{M_\infty}$, as a function of time over the initial 10-minute period. These data were then fit with **eq. 5** to obtain apparent diffusivity $D$. Though true for assumptions of constant film thickness, **eq. 5** was modified to account for the change in the coating's mean thickness obtained via optical profilometry. Further, it provided better agreement between the measured variations and those predicted by the model (see **Fig. S3** in Supplemental Information). For illustration purposes, **Figure 7** displays the drying kinetics for coatings with 4470 g/mol molecular weight resin and high RER solvent.

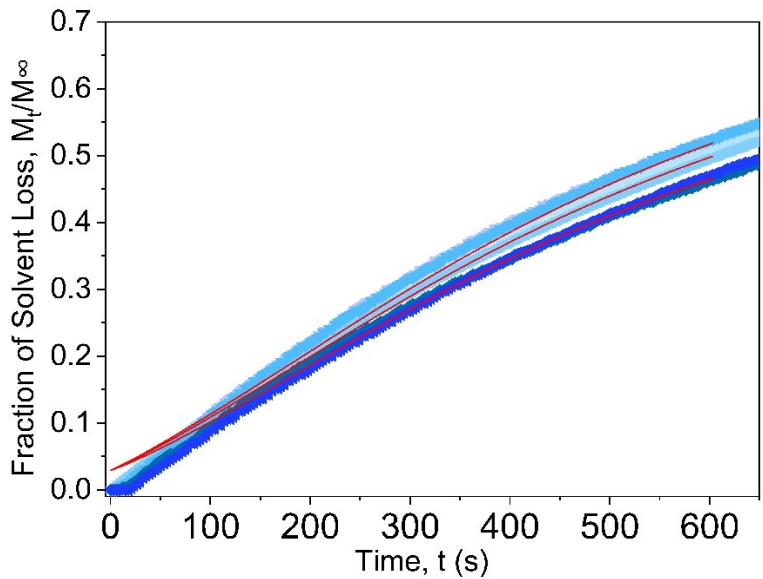

**Figure 7: Desorption curves for polymer solutions with 4470 g/mol molecular weights.** The y-axis represents the mass ratio of evaporated solvent to the total solvent mass in a 45% polymer composition by weight. The blue curves are six experimental repeats determined from gravimetry. The average fraction of solvent loss reported ~ 50% is seen by the highest value reached at the 10 minutes mark. The red lines are theoretical predictions based on a known polymer thickness. The thickness decrease was incorporated into the fitting model to improve the goodness of the fit. Drying kinetics were evaluated for the first 10 minutes following coating application. This corresponded to the flash stage during which sag measurements were conducted.

Apparent solvent diffusivity was determined in this manner for different formulations of model clear coat. **Figure 8** shows the apparent diffusivity as a function of initial mean thickness. As expected, the apparent diffusivity displayed an increase for formulations containing high RER solvent as compared to low RER solvent. As was the case with the rate constant shown above, the shift in data is attributed primarily to the difference in solvent volatility, but could also be impacted by polymer volume fraction[40,56]. High RER solvents displayed a two-fold increase in vapor pressure compared to low RER solvents. In addition, there is a 3% reduction in polymer volume fraction for the high RER systems (φ ~ 42% v/v and ~39% v/v for the low and high RER systems respectively). **Figure 8** also shows the somewhat surprising impact of initial wet film thickness. Namely, that the apparent diffusivity, for the same initial polymer weight percent, increased with increasing film thickness. This effect has been previously attributed to thicker films taking longer to harden after which solvent diffusion is hindered[64].

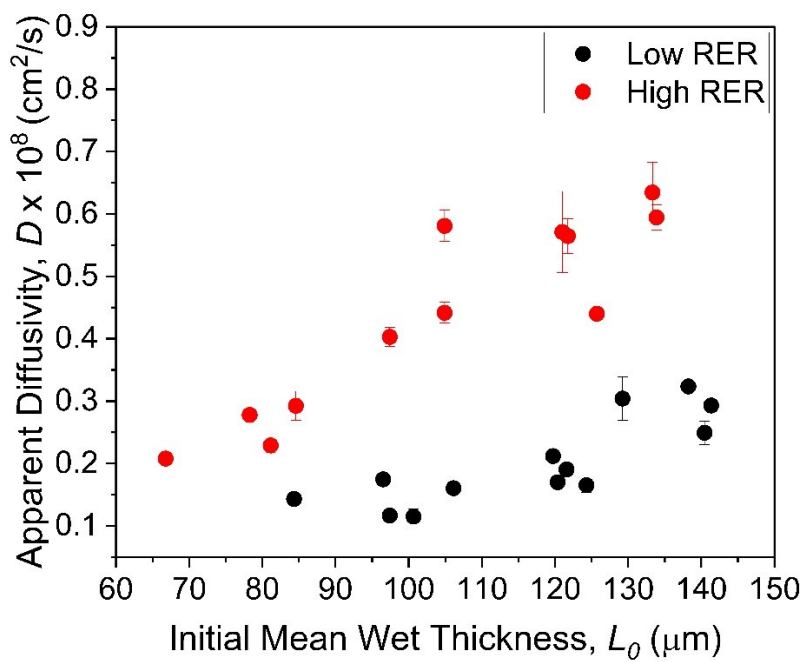

**Figure 8. Dependence of apparent diffusivity on initial mean thickness of the wet film.** The apparent diffusivity ($cm^2/s$) as a function of initial mean thickness (μm). The red circles represent formulations containing high RER solvent. The black circles are samples with low RER solvent. The error bars are standard error associated with six experimental repeats. The diffusion coefficient, for the same initial polymer weight percent, increases with increasing the film thickness. The increase in apparent diffusivity is attributed to thickness effects where thicker films take longer to dry. Further, formulations containing a high RER solvent displayed enhanced diffusivity. This shift in data is primarily due to the increase in volatility, but a ~ 3% reduction in polymer volume fraction for the high RER systems also plays contributes.

Note the important connection between molecular weight and the resulting viscosity, film thickness, and ultimately solvent mass transport through the film. As revealed with the modeling approach used for this work (see **eq. 5**), the characteristic time scale for transport of solvent through a film of thickness $L$ is $\sim L^2/D$. Experimental data was found to follow a master curve after scaling time with the quotient of the square of film thickness and apparent diffusivity (i.e. dimensionless time $Dt/L^2$). These data were subsequently averaged across all molecular weights for low and high RER solvents (see **Fig. 9**). This master curve guides our understanding on this relationship. For example, if one takes a standard level of film dryness (see **Fig. 9 inset**), for instance $M_t/M_\infty = 0.5$, this corresponds to a dimensionless time of $Dt/L^2 \approx 0.2$, and one can see that a thicker film would require longer to achieve this level of 'dryness'. Herein, our experiments showed that despite having slower evaporation as measured through $\alpha$ (see **Fig. 4**), the apparent diffusivity $D$ increased as a function of molecular weight. Recall that coatings with higher molecular weight resin had a greater viscosity and by extension a larger wet film thickness (see **Fig. 8**). This feature, namely larger thickness with molecular weight, seemingly artificially increased the apparent diffusivity obtained from the model from $\sim L^{-2}$. The diffusion length $\sqrt{Dt}$ evaluated at the conclusion of flash $t_f = 600\ s$ provides further support. The diffusion length $\sqrt{Dt_f} \approx 15\ \mu m$ indicates that the diffusion gradient has propagated only ~10% - 25% into the thickness of the film at $t_f$, suggesting that for the initial flash stage the dominant mechanism of drying is near surface evaporation. This is further reinforced by data summarized in **Figure 9**, in which experimental data only reaches a dimensionless time of $Dt/L^2 \approx 0.05$ and $Dt/L^2 \approx 0.14$ for coatings with low and high RER, respectively. These dynamics arising from the connection between molecular weight choice, viscosity, film thickness, and mass transport through the film are critically important to film processing, especially when considering the formation of defects, such as skinning and sag.

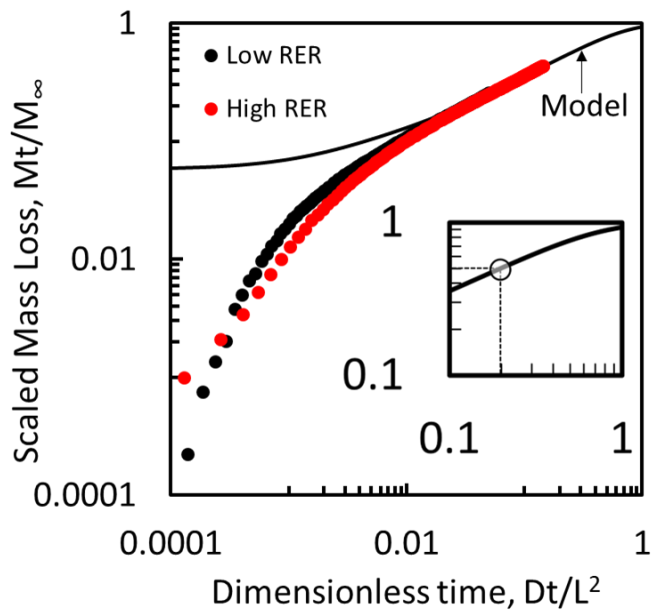

**Figure 9. Dimensionless drying curve.** Scaled mass loss is plotted as a function of dimensionless time for the model (**eq. 5**) and the average values for all molecular weights for both low (black circles) and high (red circles) RER solvents. Note the model diverges at short dimensionless times because the number of summation terms was truncated at n = 6. Drying occurs moving along the black curve to longer dimensionless times. The inset shows that are a dimensionless time value of 0.2, the expectation is that 50% of solvent will have been lost. Experiments described herein proceed to a dimensionless time value ~0.15 and ~0.05 for high and low RER solvent, respectively, indicating that most solvent has yet to evaporate in experiments mimicking the flash period of the coating process.

Sag velocity, obtained via VAIM, was measured for formulations exhibiting a systematic variation in molecular weight. The velocity of many probe particles was measured and the average velocity in the y-direction was calculated (see **Figure 1** for axis direction). Although the fundamental dependence of material flowability (sag velocity) on viscosity is known, there are process-dependent factors (i.e., film thickness) which contribute to the overall performance. For instance, increased viscosity is expected to increase flow opposition and mitigate sag[65], yet our work has thus far shown that increasing viscosity leads to a larger wet film thickness that would be thought to increase sag because of increased gravitational stress. Beyond the resin molecular weight, solvent volatility will also play a role. As coatings dry, the viscosity increases monotonically as the polymer concentration increases[66] and increasing the rate at which solvent escapes a film can lead to a more rapid change in viscosity[42]. This is expected to help mitigate sag as the viscous forces dominate[46,49,65]. While this work did not include comparative studies accounting for the influence of solvent volatility, sag velocity is expected to increase in formulations with Low RER solvent.

**Figure 10** displays the impact of resin molecular weight on sag velocity. Sag velocity decreased over time as depicted by the color gradient representative of the magnitude of velocity. The monotonic decrease in sag velocity as a function of time while drying is attributed to the combined effects of reduced film thickness and increased viscosity. Interestingly, thicker films tended to experience a reduction in sag velocity despite experiencing a higher gravitational stress and (based on our results from previous sections) a less dry state. Although our previous work suggested that thicker films would experience increased sag, those films were at fixed viscosity. Herein, the thicker films were a consequence of greater viscosity from higher molecular weight resin. This suggests that viscous forces as mediated via formulation choice may be more effective than film build when aiming to mitigate sag.

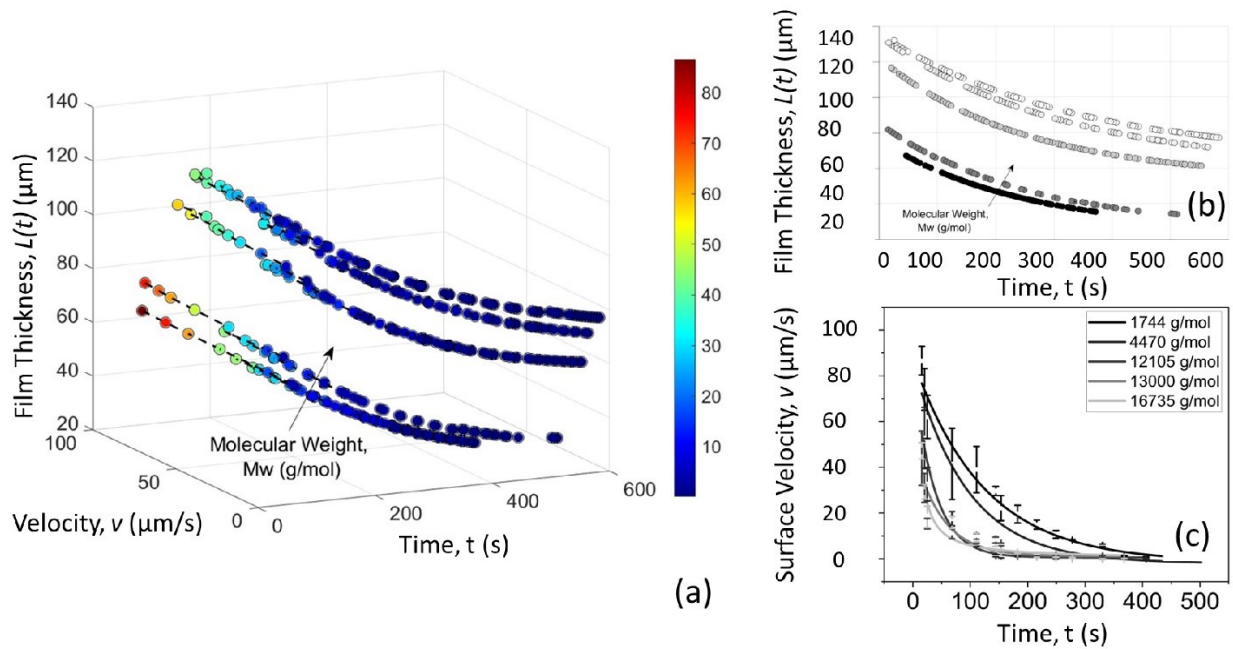

**Figure 10. Sag velocity as measured by the VAIM positioned at 5° incline.** (a) Five formulations with viscosities of 0.026, 0.033, 0.068, 0.087, 0.106, and 0.115 Pa s were evaluated with viscosity (i.e., molecular weight) increasing left to right as depicted by the arrow. (b) The y-axis is the average velocity (μm/s). The z-axis represents the average film thickness (μm) as measured with profilometry. Particle tracking was conducted for regions ~ 1/3 from the coating upstream. Each velocity point presented herein correspond to an average velocity over a 5 second period. Sag velocity is seen to decrease over time as depicted by the color gradient representative of the magnitude of velocity. Film thickness over time. (c) Increasing the formulation viscosity led to increased applied film thickness for the same blade gap (i.e., 10 mils), but decreased the surface velocity more rapidly in time.

5.  **Summary and Conclusions**

Work presented herein describes the impact of resin molecular weight on drying kinetics and sag in polymer films. The systems contained 45% polymer by weight with systematic variation in resin molecular weight and solvent volatility. These systems were drawn down at fixed clearance and subsequently measured for changes in film thickness, mass, and sag as solvent evaporated. Wet film thickness was observed to increase as a function of molecular weight, an effect that was attributed to the increase in shear viscosity for increasing molecular weight. Gravimetry experiments were used to measure the evaporative mass loss for films and was parametrized by the evaporation rate constant and residual solvent for both short (~600 s) and long (~2 hrs) times. As molecular weight increases or vapor pressure decreases, solvent tends to evaporate more slowly at short times (~10 minutes) corresponding to the flash step of a coating process. At longer times, the effect of solvent vapor pressure diminishes, but the impact of molecular weight remains such that higher molecular weight resin more effectively retains solvent. Apparent diffusivity of the solvent was also obtained to characterize the drying process. Interestingly, the apparent diffusivity of solvent was found to increase as molecular weight increased, which would be in apparent contradiction to the trend obtained from the evaporation rate constant. However, analysis of the diffusion length suggests that the concentration gradient driving evaporation only reaches a shallow depth of the film during the flash portion (~600 s) of drying, such that these data are primarily driven by near surface solvent evaporation. Inclusion of the thickness of the film in the analysis artificially enhances the apparent diffusivity. Finally, sag velocity was measured for these films and found to decrease over time as the films flow and dry. This monotonic decrease was due to reduced film thickness and increased viscosity. Surprisingly, thicker films (i.e., high resin molecular weight), displayed a more rapid reduction in sag velocity as compared to thinner films over initially lower viscosity. Hence, we conclude for these systems the effect of viscosity plays a more significant role than gravitational stress for sagging films. This work highlights the importance of formulation properties and their impact on coating processability, especially for coating applications that rely on blade or brush coating. We concluded from our work the wet film thickness is far more sensitive to viscosity than what conventionally is thought, and this dependence will extend to drying rate.

**Acknowledgements**

This work was supported by the National Science Foundation (NSF) award nos. 1931636 (C.L.W. and R.M.R.) and 1931681 (J.F.G.). We thank Professor João M. Maia (CWRU) and Professor Burcu Gurkan (CWRU) for laboratory facility use. We thank Dr. Yangming Kou (PPG) for helpful discussions.

*Supplemental Information*

Impact of Resin Molecular Weight on Drying Kinetics and Sag of Coatings

*Marola W. Issa[1], Steve V. Barancyk[2], Reza M. Rock[2], James F. Gilchrist[3], and Christopher L. Wirth[1]*

[1]Department of Chemical and Biomolecular Engineering, Case School of Engineering, Case Western Reserve University, Cleveland, Ohio 44106, United States

[2]PPG Industries, Inc. Pittsburgh, Pennsylvania 15272, United States

[3]Department of Chemical and Biomolecular Engineering, Lehigh University, Bethlehem, PA 18015, United States

**Corresponding author**

Christopher L. Wirth

Chemical and Biomolecular Engineering Department

Case School of Engineering

Case Western Reserve University

Cleveland, OH 44106

wirth@case.edu

MWI Orchid: 0000-0002-8099-0459

CLW Orchid: 0000-0003-3380-2029

Number of Figures: 6

Number of pages: 7

**List of Figures**

*Solvent desorption curve*

**Figures 1** is an example of a desorption curve with enhanced fit and the diffusion coefficient obtained with variable summation terms.

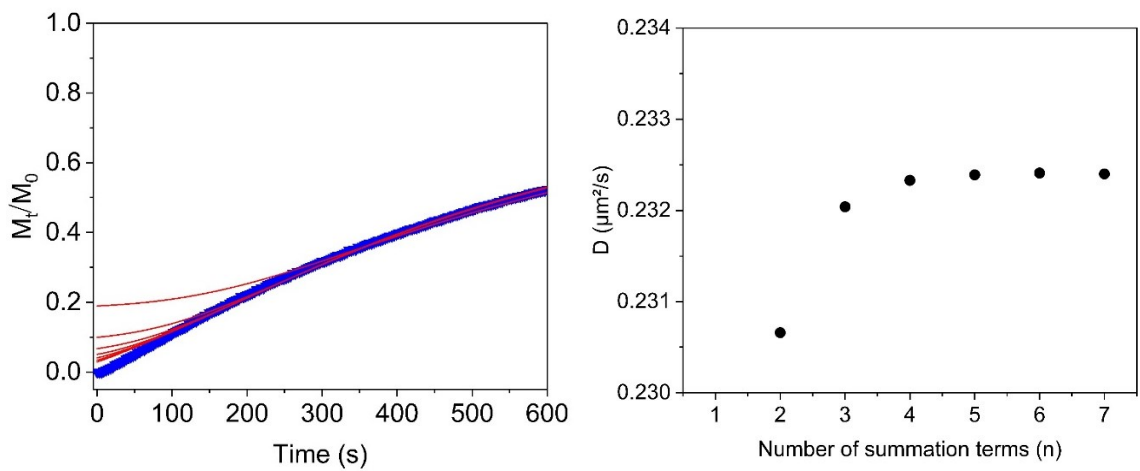

**Figure S1.** Desorption curves for a polymer solution with 4470 g/mol molecular weight. The y-axis represents the mass ratio of evaporated solvent to the total solvent mass in a 45% polymer composition by weight. The blue curve represents experimental data as determined from gravimetry. The red lines are theoretical predictions based on the number of summations used in the model. As the number of summations was increased, the fit lines are seen to approach the values measured experimentally (left). Seven summation terms were carried out for the diffusion coefficient estimate. Diffusivity values were seen to converge around n = 7 (right). The difference in two consecutive predictions for diffusivity was seen to decrease with increasing n. A value of convergence criterion of ~ 0.06% is thought acceptable and summations were terminated at seven terms.

*Sag velocity measurements via VAIM*

**Figures 2** presents three experimental repeats for a VAIM sag velocity measurement.

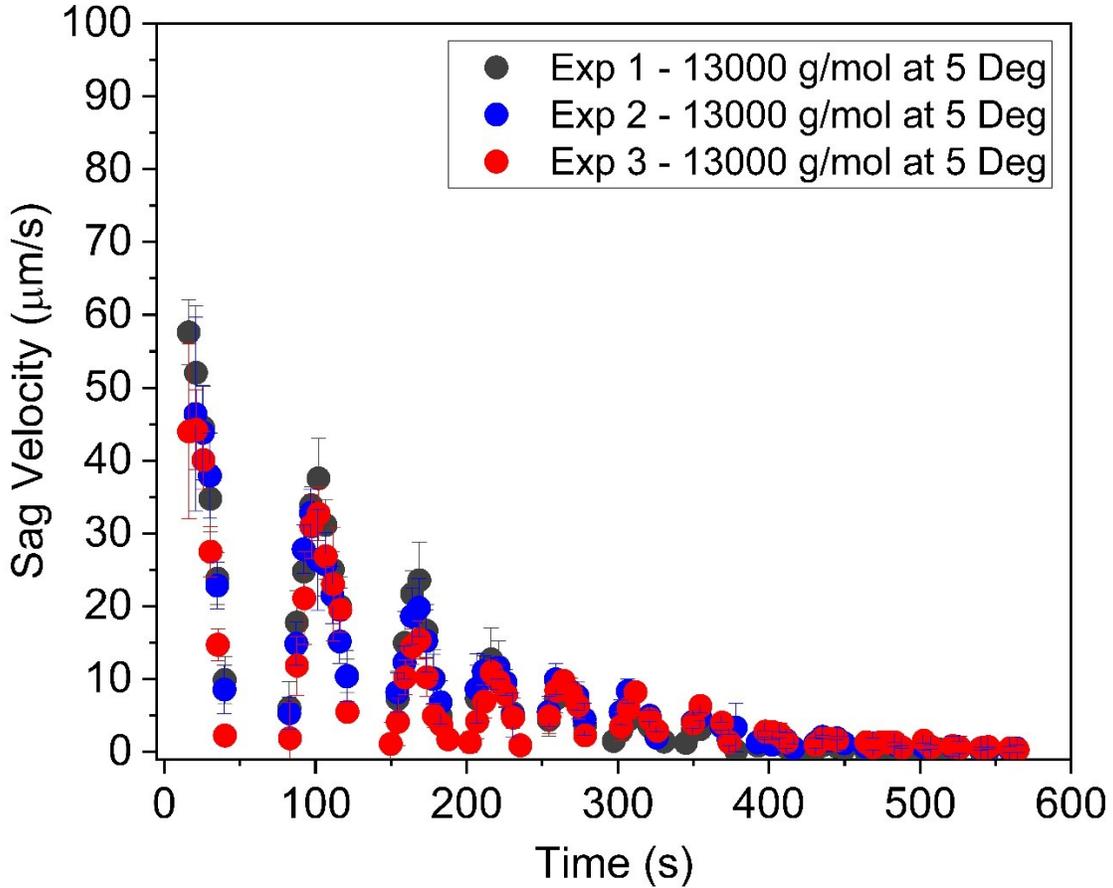

**Figure 2**: Three experimental repeats for the measured particle velocity (µm/s) over time (s). Samples with molecular weight of 13000 g/mol, containing a high RER solvent were positioned at 5° incline. The highest velocity was measured at the free surface, represented by the peaks. The velocity decreased deep into the film and reached zero at the substrate. Each peak represents one scan cycle in and out of the film starting at the free surface and then returning to the free surface. The peaks progressively decreased in size indicating a decrease in particle velocities as the film thickness decays. Error bars represent the standard deviation associated with measured velocity obtained via the particle tracking algorithm.

*Constant thickness vs. variable thickness fit evaluation*

**Figures 3** Represents an experimental sample modeled with a constant thickness assumption and variable thickness assumptions. Incorporating variable thickness as obtained from profilometry, increased the goodness of the fit.

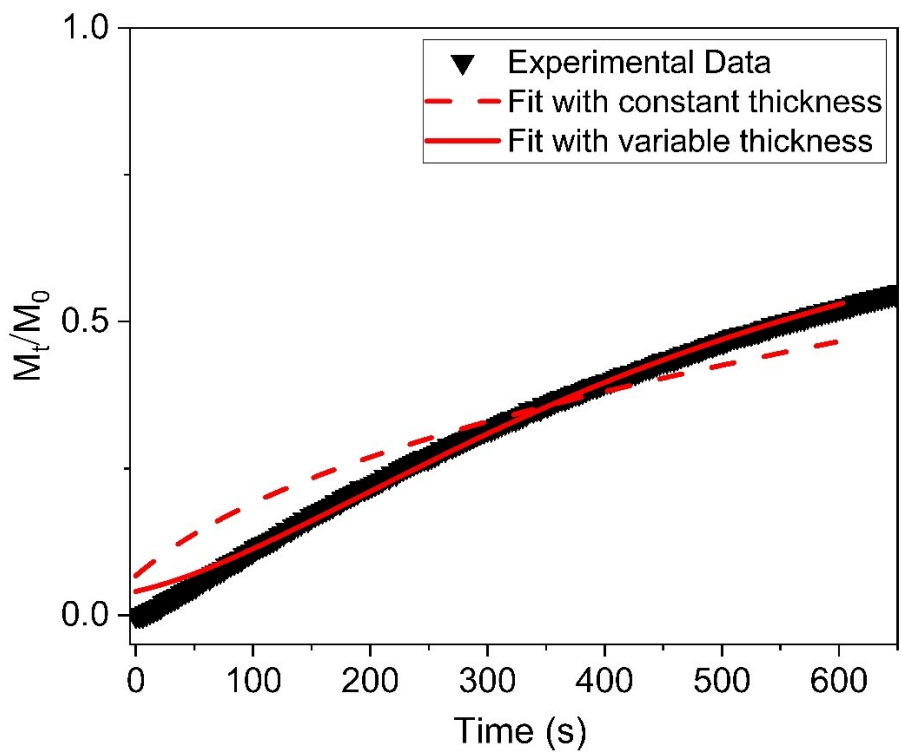

**Figure 3**: Formulation with 1129 g/mol molecular weight modeled with constant thickness assumption (dashed red line) and variable thickness assumption (solid red line). The black line represents experimental data of normalized solvent loss (x-axis) over time (y-axis). Incorporating the profilometry measurements for height decay into the desorption model increased the goodness of the fit. The R-Squared values for the fit was 0.95 and 0.99 for constant thickness and variable thickness assumptions, respectively.

*Solvent Desorption at the End of Drying*

**Figures 4** Represents three experimental desorption repeats for coatings formulated with the same molecular weight resin but variable RER solvents.

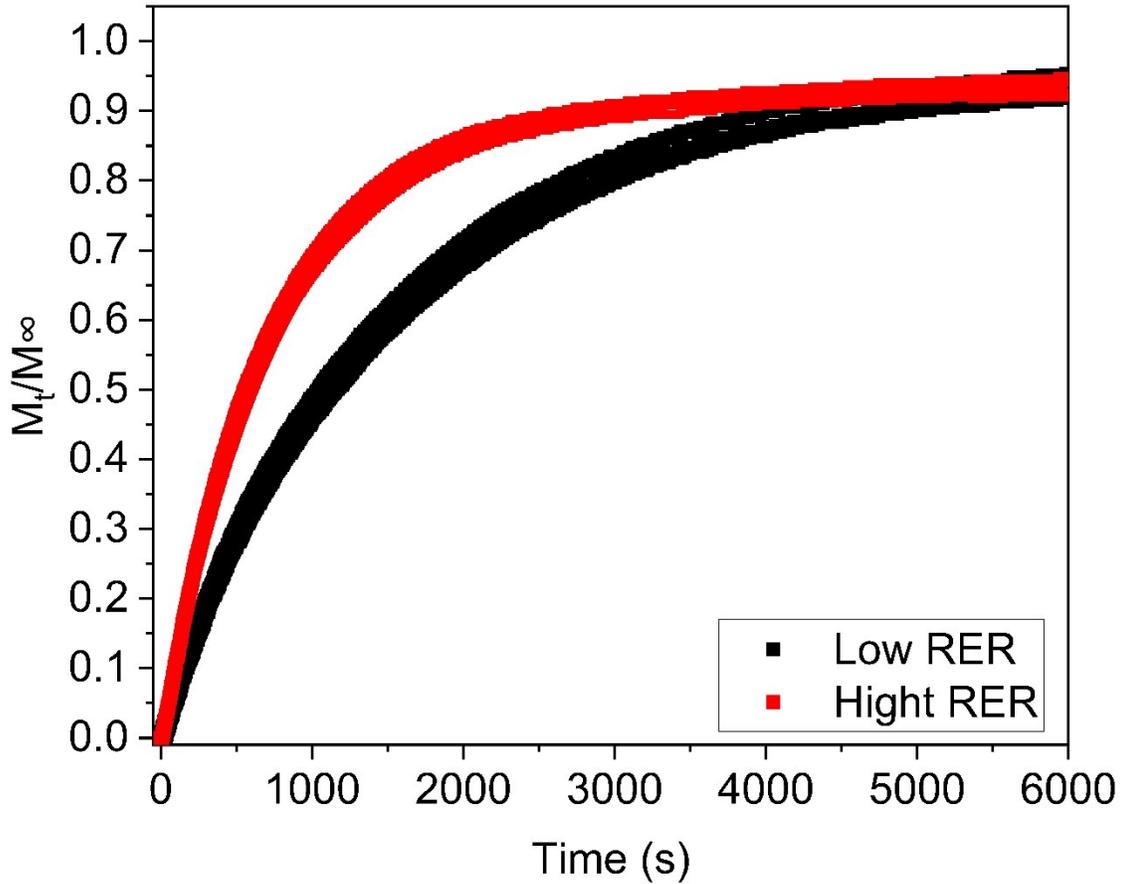

**Figures 4** Represents an experimental Desorption curve for coatings formulated with the same molecular weight resin but variable RER solvents. The residual solvents retained in the film displayed variable drying rates depending on the solvent volatility. Further, high RER solvents reach a plateau around 3000 seconds whereas low RER take longer to taper off. At the later stages of drying, the residual solvents displayed similar values, irrespective of the solvent RER.

*Residual Solvent After 10 Minutes*

**Figures 5** Represents the residual solvent remaining after 10 minutes of drying.

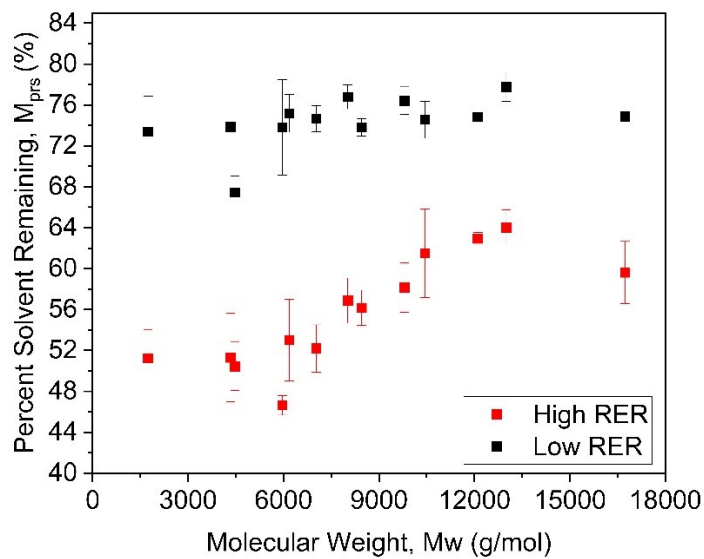

**Figure S5**: Percent residual solvent at the end of flash stage. Formulations containing high RER solvents displayed reduction in residual solvent due to increased evaporation rate. The residual solvent displayed a dependence on resin molecular weight such that an increase in molecular weight led to increased percent of trapped solvent.

*Residual Solvent at the End of Drying*

**Figures 6** Represents the residual solvent remaining after 2 hours of drying

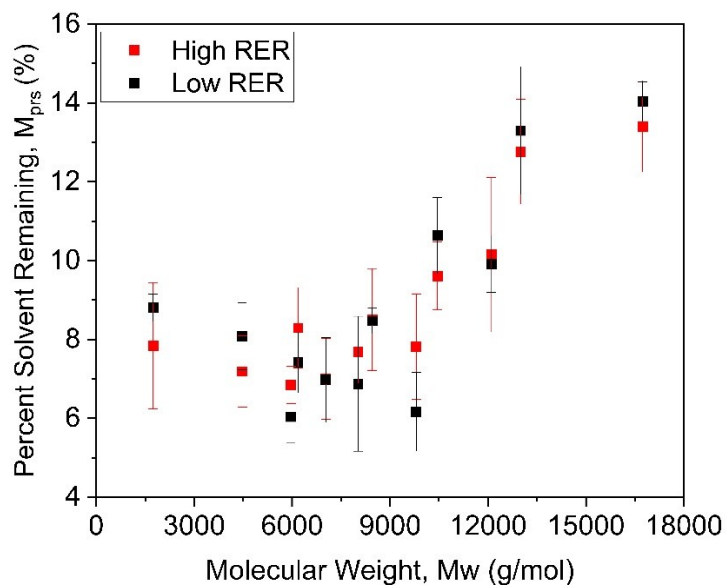

**Figure S6**: Percent residual solvent at the end of drying (i.e., 2 hours). Formulations displayed similar percentages of trapped solvent irrespective of solvent volatility. The molecular weight dependence holds true for extended evaporation times such that larger molecular weights retained larger percentages of solvent.